\documentclass[twocolumn,pre,showpacs]{revtex4}
\usepackage{amssymb}
\usepackage{amsfonts}
\usepackage{amsmath}
\usepackage{graphicx}

\begin{document}

\title{A memory-induced diffusive-superdiffusive transition: ensemble and
time-averaged observables}
\author{Adri\'{a}n A. Budini}
\affiliation{Consejo Nacional de Investigaciones Cient\'{\i}ficas y T\'{e}cnicas
(CONICET), Centro At\'{o}mico Bariloche, Avenida E. Bustillo Km 9.5, (8400)
Bariloche, Argentina, and Universidad Tecnol\'{o}gica Nacional (UTN-FRBA),
Fanny Newbery 111, (8400) Bariloche, Argentina}
\date{\today }

\begin{abstract}
The ensemble properties and time-averaged observables of a memory-induced
diffusive-superdiffusive transition are studied. The model consists in a
random walker whose transitions in a given direction depend on a weighted
linear combination of the number of both right and left previous
transitions. The diffusion process is nonstationary and its probability
develops the phenomenon of aging. Depending on the characteristic memory
parameters, the ensemble behavior may be normal, superdiffusive, or
ballistic. In contrast, the time-averaged mean squared displacement is equal
to that of a normal undriven random walk, which renders the process
non-ergodic. In addition, and similarly to Levy walks [Godec and Metzler,
Phys. Rev. Lett. \textbf{110}, 020603 (2013)], for trajectories of finite
duration the time-averaged displacement apparently become random with
properties that depend on the measurement time and also on the memory
properties. These features are related to the non-stationary power-law decay
of the transition probabilities to their stationary values. Time-averaged
response to a bias is also calculated. In contrast with Levy walks
[Froemberg and Barkai, Phys. Rev. E \textbf{87}, 030104(R) (2013)], the
response always vanishes asymptotically.
\end{abstract}

\maketitle

% 05.40.-a Fluctuation phenomena, random processes, noise, and Brownian motion
% 89.75.Da Systems obeying scaling laws
% 05.40.Fb Random walks and Levy Flights

% 87.15.Vv Diffusion (in Biomolecules)
% 02.50.Cw Probability theory
% 02.50.Ey Stochastic processes
% 02.50.-r Probability theory, stochastic processes, and statistics
% 05.70.-a Thermodynamics
% 05.70.Ln NonEquilibrium and irreversible thermodynamics
% 05.40.-a Fluctuation phenomena, random processes, noise, and Brownian motion
% 05.40.Jc Brownian motion
% 05.40.Ca Noise
% 05.40.Fb Random Walks and Levy Flights
% 05.20.Gg Classical ensemble theory
% 89.75.Da Systems obeying scaling laws
% 89.75.-k Complex systems
% 87.10.+e General theory and mathematical aspects (bilogical and medical physics)
% 87.15.Vv Diffusion (in Biomolecules)
% 87.15.Ya Fluctuations (in Biomolecules)

\section{Introduction}

Anomalous superdiffusive processes describe a wide variety of systems
arising in different disciplines such as physics and biology. Levy walks is
one of the simpler models that lead to this feature \cite%
{denisov,monti,zumo,trefan,soko,zebro}. It is a generalization of the
classical Drude model where a particle moves, in successive random
directions, with constant velocity during random periods of time. Depending
of the mean sojourn times a \textit{transition} between diffusive,
superdiffusive and ballistic behaviors is obtained \cite{denisov,zumo}.

Similarly to other anomalous diffusive processes \cite%
{burovS,burov,gbel,fuli,igor,radons,cherstvy,peters,garcia,bodrova,weron,safdari,yan,aki,maria,2017}%
, the ergodic properties of L\'{e}vy walks were recently studied \cite%
{godec,einsteinLevy,godecEinstein}. While ensemble moments are defined in a
usual way, time-averaged moments, as in single-particle tracking techniques 
\cite{maria}, are defined by a temporal moving average performed with only
one single trajectory of a given temporal length (see for example Refs. \cite%
{burovS,burov}),%
\begin{equation}
\delta _{\kappa }(t,\Delta )\equiv \frac{\int_{0}^{t-\Delta }dt^{\prime
}[X(t^{\prime }+\Delta )-X(t^{\prime })]^{\kappa }}{t-\Delta }.
\label{Definition}
\end{equation}%
Here, $X(t)$ is the walker trajectory, $\Delta $ is called the lag (or
delay) time and $\kappa =1,2.$ For normal diffusive processes (independent
random increments with a characteristic time scale), ensemble- and
time-averaged moments coincide, $\lim_{t\rightarrow \infty }\delta _{\kappa
}(t,\Delta )=\langle \lbrack X(\Delta )-X(0)]^{\kappa }\rangle ,$ situation
that defines \textit{ergodicity}. $\langle \cdots \rangle $ denotes ensemble
average. The initial condition $X(0)$ appears due to the traslational
invariance of the definition (\ref{Definition}). The so called weak
ergodicity breaking is set by the condition $\lim_{t\rightarrow \infty
}\delta _{\kappa }(t,\Delta )\neq \langle \lbrack X(\Delta )-X(0)]^{\kappa
}\rangle $ even for long $\Delta .$

For Levy walks, the behavior of the time-averaged mean square displacement [$%
\kappa =2$ in Eq. (\ref{Definition})] strongly departs from that of
subdiffusive continuous-time random walks \cite{burovS,burov} where, even at
infinite measurement times $t,$ they are intrinsically random objects. For
Levy walks in the \textit{superdiffusive regime} this randomness is absent.
Ensemble and time-averaging only differ by a constant \cite%
{godec,einsteinLevy}, effect called \textit{ultraweak ergodicity breaking} 
\cite{godec}. Nevertheless, when considering trajectories made over a finite
measurement time $(t<\infty )$ an apparent randomness emerges both in the
scaling exponents as well as in the amplitude of the time-averaged mean
square displacement. This feature can be related to trajectories where the
walker persists along a great fraction or even during the entire trajectory
with the same velocity. On the other hand, in the \textit{ballistic regime}
an intrinsic randomness similar to that of subdiffusive processes arises
when considering a shifted time-averaged moment \cite{einsteinLevy}.
Furthermore, time-averaged response to a bias \cite{nogo,akimoto} and a
corresponding generalized Einstein relation were also studied \cite%
{einsteinLevy,godecEinstein}.

The previous results were obtained from a renewal description \cite{denisov}
of the stochastic dynamics. Nevertheless, alternative underlying dynamics
may also lead to superdiffusion. For example, similar analysis were
performed by considering a deterministic diffusion model \cite{akimoto} and
also correlated random walks \cite{tejedor}. In addition, \textit{globally
correlated dynamics}, where the walker dynamics depend on the whole previous
history of transitions \cite%
{gunter,kim,kursten,kenkre,gandi,silva,coletti,baur,esguerra,katja,boyer},
also may lead to superdiffusion. Given that the ensemble properties may be
similar to those of Levy walks it become of interest to study the ergodic
properties of these strongly correlated dynamics. Added to its theoretical
interest, given an experimental situation, one may obtains specific criteria
for discriminating between different possible underlying nonequilibrium
stochastic dynamics.

In Ref. \cite{urna} we introduced a globally correlated diffusive dynamics
that leads to (ensemble) ballistic behaviors and also characterized its
time-averaged moments. Interestingly, the memory effects also lead to weak
ergodicity breaking. Asymptotically the time averaged moments becomes
intrinsically random. The first and second moments [Eq. (\ref{Definition})]
grow respectively linearly and quadratically with the lag time $\Delta .$
Nevertheless, the characteristic parameters of these dependences change
realization to realization. A fluctuation-dissipation Einstein-like relation
between the first and (a centered) second time-averaged moments for driven
and undriven dynamics respectively was also established. These features are
similar to that found in Levy walks in the ballistic regime \cite%
{einsteinLevy}. Hence, it is natural to investigate if similar results can
be obtained in a sub-ballistic regime and to explore up to which point
previous results based on renewal memoryless dynamics are intrinsic to
superdiffusive process and which are intrinsic properties of the model.

The main goal of this paper is to introduce an alternative description of
superdiffusion based on global memory effects and to study its ensemble and
time-averaged observables. The model interpolates between two previous known
dynamics: the elephant model \cite{gunter} and the urn-like model of Ref. 
\cite{urna}. The transition probability of the walker depends on a weighted
linear combination of the number of both right and left previous
transitions. Hence, jumps can be correlated or anticorrelated with the
previous history. Depending on the memory parameter values, the ensemble
behavior suffers a transition between diffusion, superdiffusion and
ballistic behaviors. The nonstationary character of the process is
explicitly shown through its correlation. In addition, the probability
evolution develops the phenomenon of aging \cite{aging,xxx,age}.

We show that in contrast to Levy walks, for infinite measurement times the
time-averaged moments strongly differ from their ensemble behavior. In fact,
they are equal to that of an undriven diffusion process. Hence, ergodicity
is broken, while an ultraweak ergodicity breaking effect only appears in the
diffusive regime. On the other hand, averages performed with finite-time
trajectories develop similar properties to that of Levy walks, that is, they
become apparently random. This feature here is related to the non-stationary
power-law decay of the transition probabilities to their stationary values.
In contrast with previous results \cite{einsteinLevy,urna}, for the studied
model we also show that time-averaged response to a bias dye out in the
asymptotic regime.

The paper is outlined as follows. In Sec. II the global correlated dynamics
is introduced. A detailed characterization of its realizations is performed.
In Sec. III the ensemble properties are presented (statistical moments,
correlation, and probability evolution). In Sec. IV the time-averaged
observables are studied. Sec. V is devoted to the Conclusions. Analytical
calculations that support the main results are presented in the Appendixes.

\section{Random walk dynamics}

The model consists in a one-dimensional walker that at successive times
perform random jumps. As in Refs. \cite{gunter,urna}, both the time and
position coordinates are discrete. Hence, in each discrete time step $%
(t\rightarrow t+\delta t)$ the walker perform a jump of length $\delta x$ to
the right or to the left. For simplicity, time and position are measured in
units of $\delta x$ and $\delta t$ respectively. The stochastic position $%
X_{t}$ at time $t$ reads%
\begin{equation}
X_{t}-X_{0}\equiv x_{t}=\sum_{t^{\prime }=1}^{t}\sigma _{t^{\prime }}.
\label{position}
\end{equation}%
Here, $X_{0}$ is the initial position, while $x_{t}$ gives the departure
with respect to it. $\sigma _{t}=\pm 1$\ is a random variable assigned to
each step. The stochastic dynamics of the variables $\{\sigma _{t^{\prime
}}\}_{t^{\prime }=1}^{t}$ is as follows. At $t=1$ (first jump or transition)
the two possible values are chosen with probability $P(\sigma _{1}=\pm
1)=q_{\pm },$ where the weights satisfy $q_{+}+q_{-}=1.$ The next values are
determinate by a conditional probability $\mathcal{T}(\sigma _{1},\cdots
\sigma _{t}|\sigma _{t+1})$ (the notation is such that $\mathcal{T}(A|B)$
gives the probability of $B$ given $A).$ This object depends on the whole
previous trajectory: $\sigma _{1},\cdots \sigma _{t}.$

The present model relies in the selection%
\begin{equation}
\mathcal{T}(\sigma _{1},\cdots \sigma _{t}|\sigma _{t+1}=\pm 1)=\frac{%
\lambda q_{\pm }+\mu t_{\pm }+(1-\mu )t_{\mp }}{t+\lambda }.
\label{Transition}
\end{equation}%
This transition probability depends on two free parameters, $\lambda $\ and $%
\mu .$ They satisfy the condition $\lambda \geq 0$ and $0\leq \mu \leq 1$\
respectively. Furthermore, $t_{+}$ and $t_{-}$ are the number of times that
the walker jumped (up to time $t$) to the right and to the left
respectively, $t=t_{+}+t_{-}.$

Depending on the memory parameters $\lambda $ and $\mu ,$\ the present model
recovers two previous studied dynamics. For $\mu =1,$ the urn-like dynamics
of Ref. \cite{urna}\ is recovered, while for $\lambda =0$ the elephant model
arises \cite{gunter} (see Refs. \cite{kim,kursten} where this model is
written in terms of the number of transitions $t_{\pm }).$

The parameter $\lambda $ allows to control the degree or \textquotedblleft
intensity\textquotedblright\ of the memory effects. In fact, in the limit $%
\lambda \rightarrow \infty $ a memoryless dynamics is recovered. On the
other hand, the role of the parameter $\mu $ is to weight the two
contributions $t_{\pm }.$ For $\mu \gtrless 1/2,$ the next jump probability
is correlated (anticorrelated) with the previous trajectory. This feature
can be lighted by using that $t=t_{+}+t_{-}$\ and $x_{t}=t_{+}-t_{-},$ which
implies%
\begin{equation}
t_{\pm }=\frac{t\pm x_{t}}{2}.  \label{tiempos}
\end{equation}%
Hence, Eq.(\ref{Transition}) can be rewritten as%
\begin{equation}
\mathcal{T}(\sigma _{1},\cdots \sigma _{t}|\sigma _{t+1}=\pm 1)=\frac{%
\lambda q_{\pm }+(t\pm \alpha x_{t})/2}{t+\lambda },  \label{transitionX}
\end{equation}%
where for shortening the expression we defined the parameter $\alpha \equiv
2\mu -1.$ The previous equation say us that the right-left transitions
depend on the position $x_{t}$ of the walker. The influence of this
dependence becomes evident by considering the regime in which $t\gg \lambda
, $%
\begin{equation}
\mathcal{T}(\sigma _{1},\cdots \sigma _{t}|\sigma _{t+1}=\pm 1)\simeq \frac{1%
}{2}\Big{(}1\pm \alpha \frac{x_{t}}{t}\Big{)},  \label{TranAsimp}
\end{equation}%
where the condition $|x_{t}|\leq t$ guarantees positivity. Therefore, we
notice that when $\mu \gtrless 1/2$ $(\alpha \gtrless 0),$ for increasing
(decreasing) $x_{t}$ the next jump at $t+1$ occurs with more probability in
the positive (negative) direction than in negative (positive) direction.
While this dependence introduce a strong correlation along the trajectory,
it is possible to demonstrate that the (random) times during which the
system walks in the same direction (sojourn times) have a finite average,
that is, they are not characterized by power-law statistical behaviors (see
Appendix \ref{TIMES}).%
%figura1%figura%figura%figura%figurav%figura%figura%figura%figura%figura%figura%figura%figura%figura%figurav%figura%figura%figura%figura%figura
%figura%figura%figura%figura%figurav%figura%figura%figura%figura%figura%figura%figura%figura%figura%figurav%figura%figura%figura%figura%figura
\begin{figure}[tbp]
\includegraphics[bb=25 315 735 1145,angle=0,width=8.5cm]{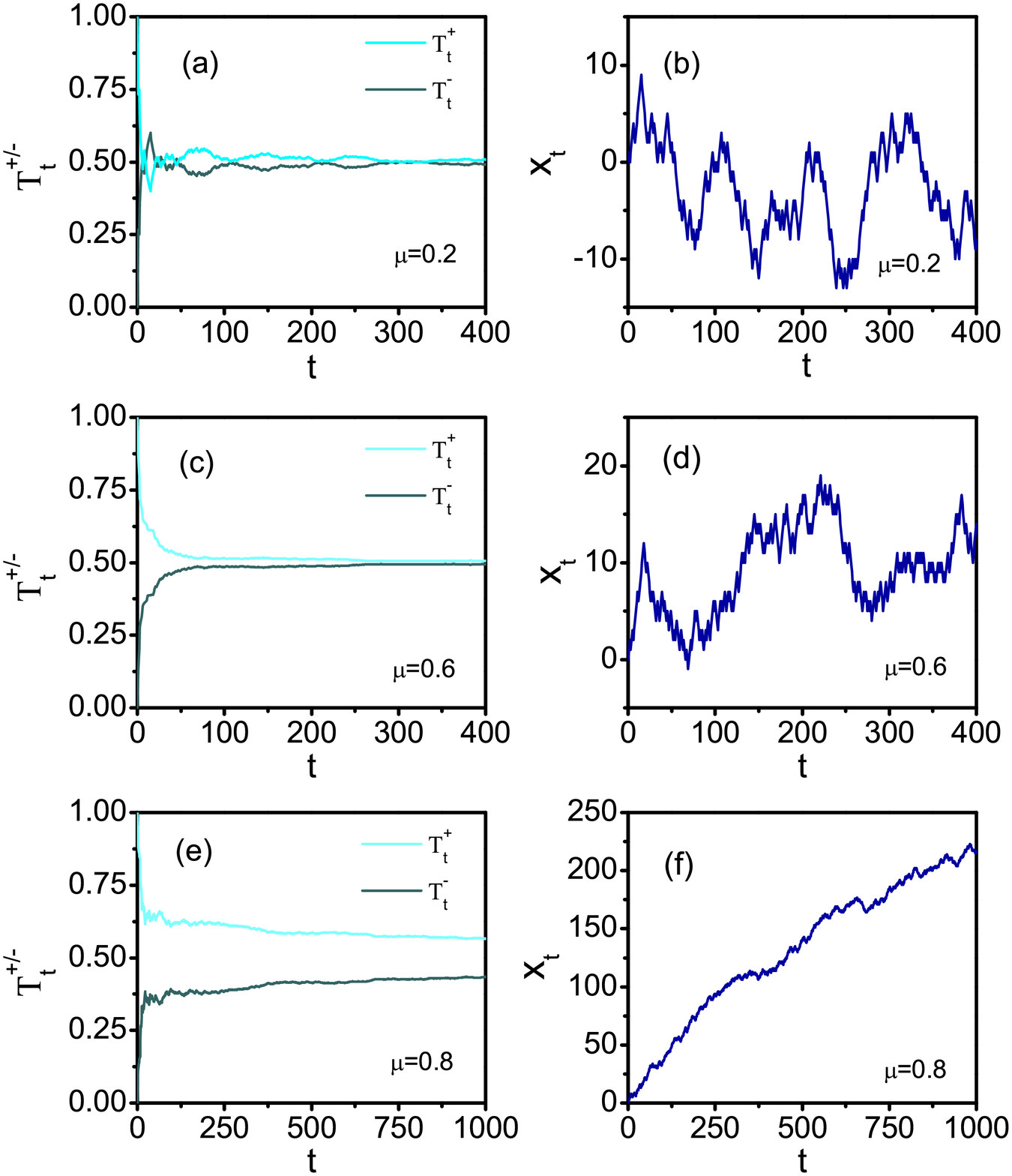}
\caption{Different realizations of the transition probabilities $\mathcal{T}%
_{t}^{\pm }=\mathcal{T}(\protect\sigma _{1},\cdots \protect\sigma _{t}|%
\protect\sigma _{t+1}=\pm 1)$ [Eq. (\protect\ref{Transition})] jointly with
the corresponding centered walker trajectory $x_{t}$ [Eq. (\protect\ref%
{position})] as a function of time. In all cases, $\protect\lambda =2$ and $%
q_{+}=1,$ $q_{-}=0.$ The value of $\protect\mu $ is indicated in each plot.}
\end{figure}
%figura%figura%figura%figura%figurav%figura%figura%figura%figura%figura%figura%figura%figura%figura%figurav%figura%figura%figura%figura%figura
%figura%figura%figura%figura%figurav%figura%figura%figura%figura%figura%figura%figura%figura%figura%figurav%figura%figura%figura%figura%figura

\subsection{Stationary transition probabilities}

For $\mu =1,$ it is known that in the asymptotic regime $(t\gg \lambda )$
the transition probability [Eq. (\ref{Transition})] becomes a random
variable characterized by a Beta probability density \cite{urna}. On the
other hand, for $\lambda =0$ (elephant model) the previous randomness is
absent. These results were demonstrated in Ref. \cite{adrian} by analyzing
weak ergodicity breaking in globally correlated finite systems, which in
contrast to diffusive ones are endowed with a stationary state \cite%
{rebenStationary}.

From the previous limiting behaviors, it becomes of interest to determine
the stationary transition probabilities for the present model. Denoting $%
\mathcal{T}_{t}^{\pm }\equiv \mathcal{T}(\sigma _{1},\cdots \sigma
_{t}|\sigma _{t+1}=\pm 1),$ these quantities are $\mathcal{T}_{\infty }^{\pm
}\equiv \lim_{t\rightarrow \infty }\mathcal{T}_{t}^{\pm },$ which from Eq. (%
\ref{Transition}) can be written as 
\begin{subequations}
\begin{eqnarray}
\mathcal{T}_{\infty }^{\pm } &=&\lim_{t\rightarrow \infty }\frac{\lambda
q_{\pm }+\mu t_{\pm }+(1-\mu )t_{\mp }}{t+\lambda }, \\
&=&\mu \lim_{t\rightarrow \infty }\frac{t_{\pm }}{t}+(1-\mu
)\lim_{t\rightarrow \infty }\frac{t_{\mp }}{t}.
\end{eqnarray}%
In this expression, $\lim_{t\rightarrow \infty }t_{\pm }/t$ are the
asymptotic fraction of right-left transitions. Consistently, these values
must to coincide with the asymptotic transition probabilities, that is, $%
\mathcal{T}_{\infty }^{\pm }=\lim_{t\rightarrow \infty }t_{\pm }/t.$ Hence,
the previous equation leads to 
\end{subequations}
\begin{equation}
\mathcal{T}_{\infty }^{+}=\mathcal{T}_{\infty }^{-}=\frac{1}{2},\ \ \ \ \ \
\ \ 0\leq \mu <1,  \label{UnMedio}
\end{equation}%
while for $\mu =1$ none condition for $\mathcal{T}_{\infty }^{\pm }$ is
obtained. In fact, in this case $\mathcal{T}_{\infty }^{\pm }=f_{\pm },$
where $f_{\pm }$\ are Beta random variables whose probability density is $%
\mathcal{P}(f_{\pm })=\mathcal{N}^{-1}f_{+}^{\lambda _{+}-1}\ f_{-}^{\lambda
_{-}-1},$ with $\mathcal{N}=\Gamma (\lambda _{+})\Gamma (\lambda
_{-})/\Gamma (\lambda )$ \cite{urna,budini}, where $\lambda _{\pm }\equiv
\lambda q_{\pm }.$ Notice that Eq. (\ref{UnMedio}) is equivalent to the
probability transitions of a memoryless unbiased discrete diffusion process.
This result is independent of the parameter $\lambda $ and the weights $%
q_{\pm }.$

In order to check the result (\ref{UnMedio}) in Fig.~1, we plot the time
dependence of the transition probabilities for different values of $\mu ,$
jointly with the corresponding realizations of the centered walker
displacement $x_{t}$ [Eq. (\ref{position})]. For $\mu =0.2$ and $\mu =0.6$
the transition probabilities converges in a fast way to $1/2.$ Consistently,
the realization of $x_{t}$ looks like a standard diffusion process. On the
other hand, for $\mu =0.8$ the same asymptotic values $(1/2)$ are attained.
Nevertheless, the convergence is much slower. In fact, in a small time scale 
$\mathcal{T}_{t}^{\pm }$ seem to attain stationary random values, property
characteristic of the case $\mu =1$ (see Fig.~1(d) in Ref. \cite{adrian}).
Due to this feature, the gap between $\mathcal{T}_{t}^{+}$ and $\mathcal{T}%
_{t}^{-}$ drives the walker in one single direction, property clearly seen
in the trajectory of $x_{t}.$

\subsection{Power-law convergence to the stationary transition probabilities}

The plots of Fig.~1 are consistent with the asymptotic values defined by Eq.
(\ref{UnMedio}). On the other hand, the rate at which these values are
attained strongly depend of $\mu .$ In order to characterize this property
we introduce the difference $\delta \mathcal{T}_{t}$ between the transition
probabilities%
\begin{equation}
\delta \mathcal{T}_{t}\equiv \mathcal{T}_{t}^{+}-\mathcal{T}_{t}^{-}=\frac{%
\lambda \delta q+\alpha (t_{+}-t_{-})}{t+\lambda },  \label{DeltaTran}
\end{equation}%
result that follows straightforwardly from Eq. (\ref{Transition}), where $%
\delta q\equiv (q_{+}-q_{-}),$ $\delta \mathcal{T}_{t=0}=\delta q,$ and as
before $\alpha =(2\mu -1).$ Notice that $\delta \mathcal{T}_{t}$ can be read
as the \textit{instantaneous drift} felt by the walker.%
%figura1%figura%figura%figura%figurav%figura%figura%figura%figura%figura%figura%figura%figura%figura%figurav%figura%figura%figura%figura%figura
%figura%figura%figura%figura%figurav%figura%figura%figura%figura%figura%figura%figura%figura%figura%figurav%figura%figura%figura%figura%figura
\begin{figure}[tbp]
\includegraphics[bb=35 590 725 1145,angle=0,width=8.5cm]{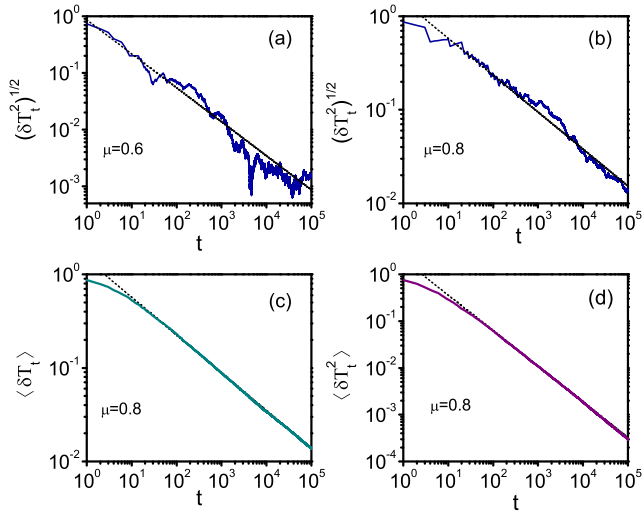}
\caption{Stochastic trajectories of the transition probability difference $%
\protect\sqrt{\protect\delta \mathcal{T}_{t}^{2}}$ for (a) $\protect\mu =0.6$
and (b) $\protect\mu =0.8.$ In (c) and (d), the full lines correspond to the
ensemble averages $\langle \protect\delta \mathcal{T}_{t}\rangle $ and $%
\langle \protect\delta \mathcal{T}_{t}^{2}\rangle $ respectively. The
average is performed with 200 realizations and $\protect\mu =0.8.$ The
dotted lines are power law fits, $\langle \protect\delta \mathcal{T}%
_{t}\rangle \simeq 1.44/t^{0.40}$ and $\langle \protect\delta \mathcal{T}%
_{t}^{2}\rangle \simeq 2.07/t^{0.76}.$ In all cases we take $\protect\lambda %
=2$ and $q_{+}=1,$ $q_{-}=0.$}
\end{figure}
%figura%figura%figura%figura%figurav%figura%figura%figura%figura%figura%figura%figura%figura%figura%figurav%figura%figura%figura%figura%figura
%figura%figura%figura%figura%figurav%figura%figura%figura%figura%figura%figura%figura%figura%figura%figurav%figura%figura%figura%figura%figura

In Figs. 2(a) and 2(b) we plot $\sqrt{\delta \mathcal{T}_{t}^{2}}.$ We find
that, after a initial transient, independently of the parameter values of
the model, $\sqrt{\delta \mathcal{T}_{t}^{2}}\approx c/t^{\beta },$ where $c$
and $\beta $\ change in each realization. As shown by the figures, this
behavior is valid over many decades of time. For $\mu <1/2$ (not shown) the
signal $\sqrt{\delta \mathcal{T}_{t}^{2}}$ become more noisy [see Fig.~1(a)]
but a power-law decay behavior is also present.

In order to characterize the previous decay behaviors in Fig.~2(c) and 2(d)
we plot $\langle \delta \mathcal{T}_{t}\rangle $ and $\langle \delta 
\mathcal{T}_{t}^{2}\rangle $ for the same value of $\mu .$ $\langle \cdots
\rangle $ denotes average over an ensemble of realizations. For both
quantities we find that asymptotically a power-law fitting always apply%
\begin{equation}
\langle \delta \mathcal{T}_{t}\rangle \simeq \frac{c_{1}}{t^{\beta _{1}}},\
\ \ \ \ \ \ \ \ \ \ \ \ \langle \delta \mathcal{T}_{t}^{2}\rangle \simeq 
\frac{c_{2}}{t^{\beta _{2}}}.  \label{BetasDefinition}
\end{equation}%
The time scale where this fitting start to be valid strongly depend on $\mu
. $ Nevertheless, when achieved, we found that the scaling exponents $\beta
_{1}$ and and $\beta _{2}$ only depend on the parameter $\mu .$

\subsection{Memory-induced transition}

A (memory-induced) transition is found when analyzing the dependences of the
scaling exponents$\ \beta _{1}$ and $\beta _{2}$ with the parameter $\mu .$
They can be determinate in a numerical way [see Figs.~2(c) and 2(d)],
results shown in Fig.~3. The scaling exponents can be fit as%
\begin{equation}
\beta _{1}=2(1-\mu ),  \label{Beta1}
\end{equation}%
while for the second moment as$\ (\mu \neq 1/2,$ $\mu \neq 1)$%
\begin{equation}
\beta _{2}=\left\{ 
\begin{array}{ccc}
4(1-\mu ) & \ \ if\ \  & 3/4\leq \mu <1 \\ 
&  &  \\ 
1 & \ \ if\ \  & 0\leq \mu \leq 3/4%
\end{array}%
\right. .  \label{fiteo}
\end{equation}%
While $\beta _{1}$ presents a monotonous linear behavior, the dependence of $%
\beta _{2}$ with $\mu $ suffer a transition at $\mu =3/4.$ This is an
intrinsic property of the correlation mechanism defined by the transition
probability (\ref{Transition}), which in turn is independent of the
parameters $\lambda $ and $q_{\pm }.$ 
%figura1%figura%figura%figura%figurav%figura%figura%figura%figura%figura%figura%figura%figura%figura%figurav%figura%figura%figura%figura%figura
%figura%figura%figura%figura%figurav%figura%figura%figura%figura%figura%figura%figura%figura%figura%figurav%figura%figura%figura%figura%figura
\begin{figure}[tbp]
\includegraphics[bb=45 866 730 1145,angle=0,width=8.5cm]{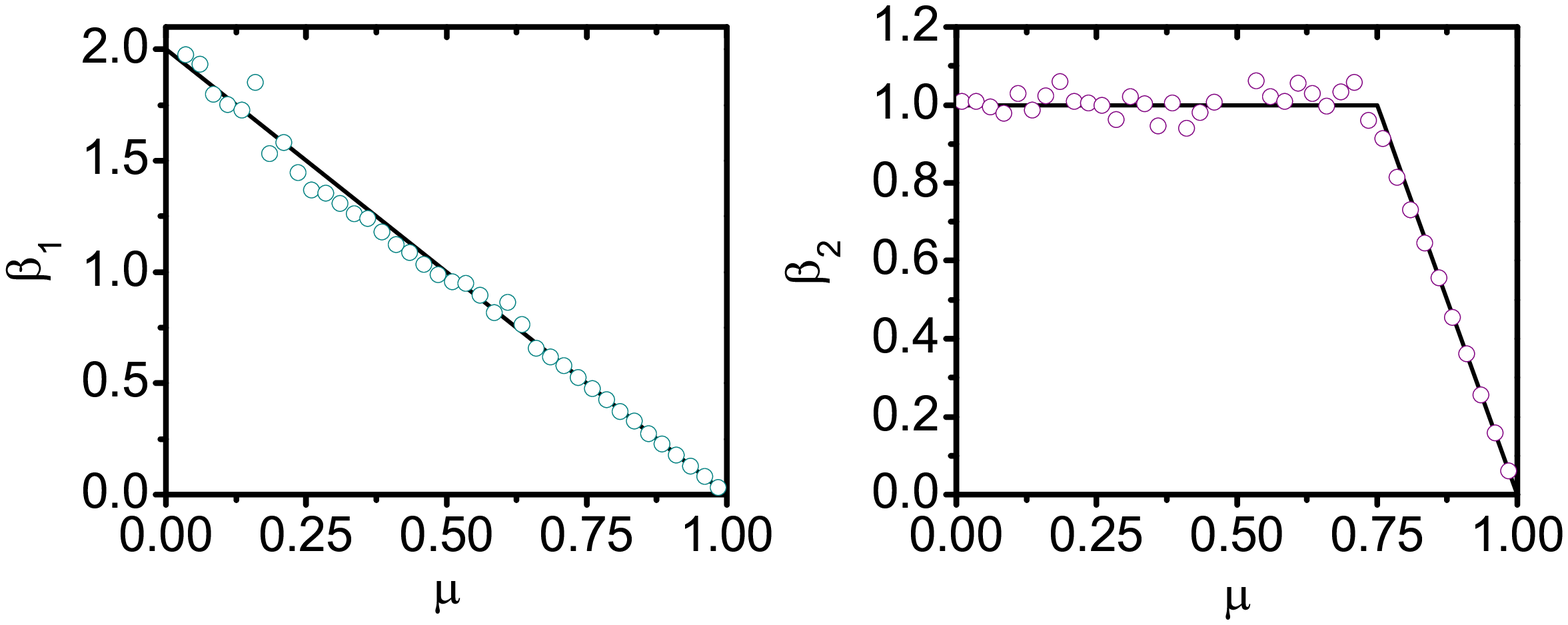}
\caption{Dependence with the parameter $\protect\mu $\ of the scaling
exponents (a) $\protect\beta _{1}$ and (b) $\protect\beta _{2}$
corresponding to the ensemble averages $\langle \protect\delta \mathcal{T}%
_{t}\rangle $ and $\langle \protect\delta \mathcal{T}_{t}^{2}\rangle $
respectively, Eq. (\protect\ref{BetasDefinition}). The circles were obtained
from numerical simulations [see Figs. (2c) and (2d)], while the full lines
gives their fitting, Eqs. (\protect\ref{Beta1}) and (\protect\ref{fiteo}).}
\end{figure}
%figura%figura%figura%figura%figurav%figura%figura%figura%figura%figura%figura%figura%figura%figura%figurav%figura%figura%figura%figura%figura
%figura%figura%figura%figura%figurav%figura%figura%figura%figura%figura%figura%figura%figura%figura%figurav%figura%figura%figura%figura%figura

For $\beta _{2}$ two values of $\mu $ are not described by Eq. (\ref{fiteo}%
). First, for $\mu =1,$ $\delta \mathcal{T}_{t}$ converges to $%
\lim_{t\rightarrow \infty }\delta \mathcal{T}_{t}=f_{+}-f_{-},$ where $%
f_{\pm }$ are Beta random variables that realization to realization satisfy $%
f_{\pm }\neq 1/2$ \cite{urna}. Therefore, in this case the exponent $\beta
_{2}$ [Eq. (\ref{BetasDefinition})] losses its meaning.

For $\mu =1/2,$ from Eq. (\ref{DeltaTran}) [with $\alpha =0]$ it follows the
deterministic behavior $\delta \mathcal{T}_{t}=\lambda \delta q/(t+\lambda
), $ leading to $\beta _{1}=1$ and $\beta _{2}=2.$ Hence, this value of $%
\beta _{2}$ is not covered by the fitting (\ref{fiteo}). Numerically, we
checked that this is the only exception. Consistently, we found that around
this point $(\mu \simeq 1/2)$ the $1/t^{\beta _{2}}$ power-law decay of $%
\langle \delta \mathcal{T}_{t}^{2}\rangle $ occurs at higher times. These
properties and results are supported by analytical calculations presented in
next sections.

\subsection{Relation between average transition fluctuations and walker
ensemble properties}

The memory-induced transition defined from the asymptotic decay of $\langle
\delta \mathcal{T}_{t}^{2}\rangle $ is analytically demonstrated in the next
section by studying the ensemble properties of the random walker
trajectories. In fact, each trajectory of $\delta \mathcal{T}_{t}$ [Eq. (\ref%
{DeltaTran})], given that $x_{t}=t_{+}-t_{-},$ can be written as $\delta 
\mathcal{T}_{t}=(\lambda \delta q+\alpha x_{t})/(t+\lambda ).$ Hence,%
\begin{equation}
\langle \delta \mathcal{T}_{t}\rangle =\frac{\lambda \delta q+\alpha \langle
x_{t}\rangle }{t+\lambda },  \label{DeltaTAverage}
\end{equation}%
while the second moment becomes%
\begin{equation}
\langle \delta \mathcal{T}_{t}^{2}\rangle =\langle \delta \mathcal{T}%
_{t}\rangle ^{2}+\frac{\alpha ^{2}[\langle x_{t}^{2}\rangle -\langle
x_{t}\rangle ^{2}]}{(t+\lambda )^{2}}.  \label{Delta2Average}
\end{equation}%
Furthermore, defining $\ \widetilde{\delta \mathcal{T}_{t}}=\delta \mathcal{T%
}_{t}-\langle \delta \mathcal{T}_{t}\rangle ,$ it follows that%
\begin{equation}
\langle \widetilde{\delta \mathcal{T}}_{t+\tau }\widetilde{\delta \mathcal{T}%
_{t}}\rangle =\frac{\alpha ^{2}[\langle x_{t+\tau }x_{t}\rangle -\langle
x_{t+\tau }\rangle \langle x_{t}\rangle ]}{(t+\tau +\lambda )(t+\lambda )}.
\end{equation}

The previous relations demonstrate that the statistical properties of $%
\delta \mathcal{T}_{t}$ and those of $x_{t}\ $can be put in one-to-one
correspondence. This relation is also valid for the variable $\sigma _{t},$
which can be read as the \textquotedblleft walker
velocity.\textquotedblright\ Given that $\mathcal{T}_{t}^{\pm }$ gives the
probability for $\sigma _{t+1}=\pm 1,$ it follows%
\begin{equation}
\langle \sigma _{t+1}\rangle =\langle \delta \mathcal{T}_{t}\rangle ,\ \ \ \
\ \ \ \ \ \ \ \ \ \ \ \ \langle \sigma _{t+1}^{2}\rangle =1.
\end{equation}%
On the other hand, defining the centered velocity $\tilde{\sigma}_{t}\equiv
\sigma _{t}-\langle \sigma _{t}\rangle ,$ its correlation reads%
\begin{equation}
\langle \widetilde{\sigma }_{t+\tau +1}\widetilde{\sigma }_{t+1}\rangle
=\langle \widetilde{\delta \mathcal{T}}_{t+\tau }\widetilde{\delta \mathcal{T%
}_{t}}\rangle .
\end{equation}%
Therefore, the properties of $\sigma _{t}$ can also be determine from the
ensemble behavior of $x_{t}.$

\section{Ensemble properties}

In this section we study the ensemble properties (moments and correlation)
of the random walk defined by Eq. (\ref{Transition}). They not only
determine the moments of the transition probability difference $\delta 
\mathcal{T}_{t}$, but also set the behavior of the time-averaged observables
(next section). The probability of finding the walker at a given time is
also obtained.

The walker statistical moments can be obtained in an exact analytical way by
introducing the double characteristic function%
\begin{equation}
Q(k_{1},t;k_{2},\tau )\equiv \langle \exp [i(k_{1}x_{t}+k_{2}x_{t+\tau
})]\rangle .  \label{doble}
\end{equation}%
In Appendix \ref{QFourier} we obtain an explicit recurrence relation for
this object. As usual, recursive relations for the moments follows by
differentiation with respect to $k_{1}$\ and $k_{2}.$ Below, we also provide
the corresponding exact solutions. Numerical simulations support the
following analytical results.

\subsection{First moment}

For the \textit{first moment,} it follows the recursive relation%
\begin{equation}
\langle x_{t+1}\rangle =\langle x_{t}\rangle \left[ 1+\frac{\alpha }{%
t+\lambda }\right] +\frac{\lambda }{t+\lambda }\delta q,
\end{equation}%
$\alpha =2\mu -1.$ Notice that for $\lambda \neq 0,$ the factor $\delta
q=q_{+}-q_{-}$ can be read as an external bias that drift the average
dynamics.

The solution of the previous equation is%
\begin{equation}
\langle x_{t}\rangle =\frac{\delta q}{\alpha }\Big{[}\frac{\Gamma (\lambda
+1)}{\Gamma (\alpha +\lambda )}\frac{\Gamma (\alpha +\lambda +t)}{\Gamma
(\lambda +t)}-\lambda \Big{]},  \label{First}
\end{equation}%
where $\Gamma (z)$\ is the Gamma function. For $\mu =1/2,$ that is $\alpha
=0,$ it follows $\langle x_{t}\rangle =\delta q\lambda \lbrack \psi (\lambda
+t)-\psi (\lambda )],$ where the digamma function is defined as $\psi
(z)=(d/dz)\ln [\Gamma (z)].$ At $\mu =1,$ $\langle x_{t}\rangle =\delta qt.$

In the long time limit, $t\gg \lambda ,$ by using the approximation $\Gamma
(z+v)/\Gamma (z)\simeq z^{v}$ valid for $z\rightarrow \infty ,$ from Eq. (%
\ref{First}) we get the asymptotic behaviors%
\begin{equation}
\langle x_{t}\rangle \approx \left\{ 
\begin{array}{ccc}
\frac{\delta q}{(2\mu -1)}\frac{\Gamma (\lambda +1)}{\Gamma (2\mu -1+\lambda
)}t^{(2\mu -1)} &  & \mu >1/2, \\ 
&  &  \\ 
\delta q\lambda \ln (t) &  & \mu =1/2, \\ 
&  &  \\ 
\frac{\delta q\lambda }{(1-2\mu )} &  & \mu <1/2.%
\end{array}%
\right.  \label{xmedio}
\end{equation}%
By taking into account these asymptotic behaviors, from Eqs. (\ref%
{DeltaTAverage}) and (\ref{First}) it is possible to confirm the fitting for 
$\beta _{1}$ given by Eq. (\ref{Beta1}).

For $\delta q=q_{+}-q_{-}\neq 0$ the first moment grows indefinitely with
time when $\mu \geq 1/2$ and saturates to a constant value when $\mu <1/2.$
In order to check these properties, in Fig.~4(a) we plot $\langle
x_{t}\rangle $ obtained from an ensemble of stochastic realizations such as
those shown in Fig. 1. Theoretical and numerical results are
indistinguishable in the scale of the plots.

The different behaviors shown in Fig.~4(a) are a consequence of the
power-law decay of $\delta \mathcal{T}_{t}$ to its stationary value, $%
\lim_{t\rightarrow \infty }\delta \mathcal{T}_{t}=0$ [see Eq. (\ref{UnMedio}%
)]. In fact, from the dependence of the exponent $\beta _{1}$ with parameter 
$\mu $ [Eq. (\ref{Beta1})] and the relation between $\langle \delta \mathcal{%
T}_{t}\rangle $ and $\langle x_{t}\rangle $ [Eq. (\ref{DeltaTAverage})],
which can be rewritten as $\alpha \langle x_{t}\rangle =\langle \delta 
\mathcal{T}_{t}\rangle (t+\lambda )-\lambda \delta q,$ it follows that the
first moment grows in time only for $\mu \geq 1/2.$

%figura1%figura%figura%figura%figurav%figura%figura%figura%figura%figura%figura%figura%figura%figura%figurav%figura%figura%figura%figura%figura
%figura%figura%figura%figura%figurav%figura%figura%figura%figura%figura%figura%figura%figura%figura%figurav%figura%figura%figura%figura%figura

\begin{figure}[tbp]
\includegraphics[bb=45 866 730 1145,angle=0,width=8.5cm]{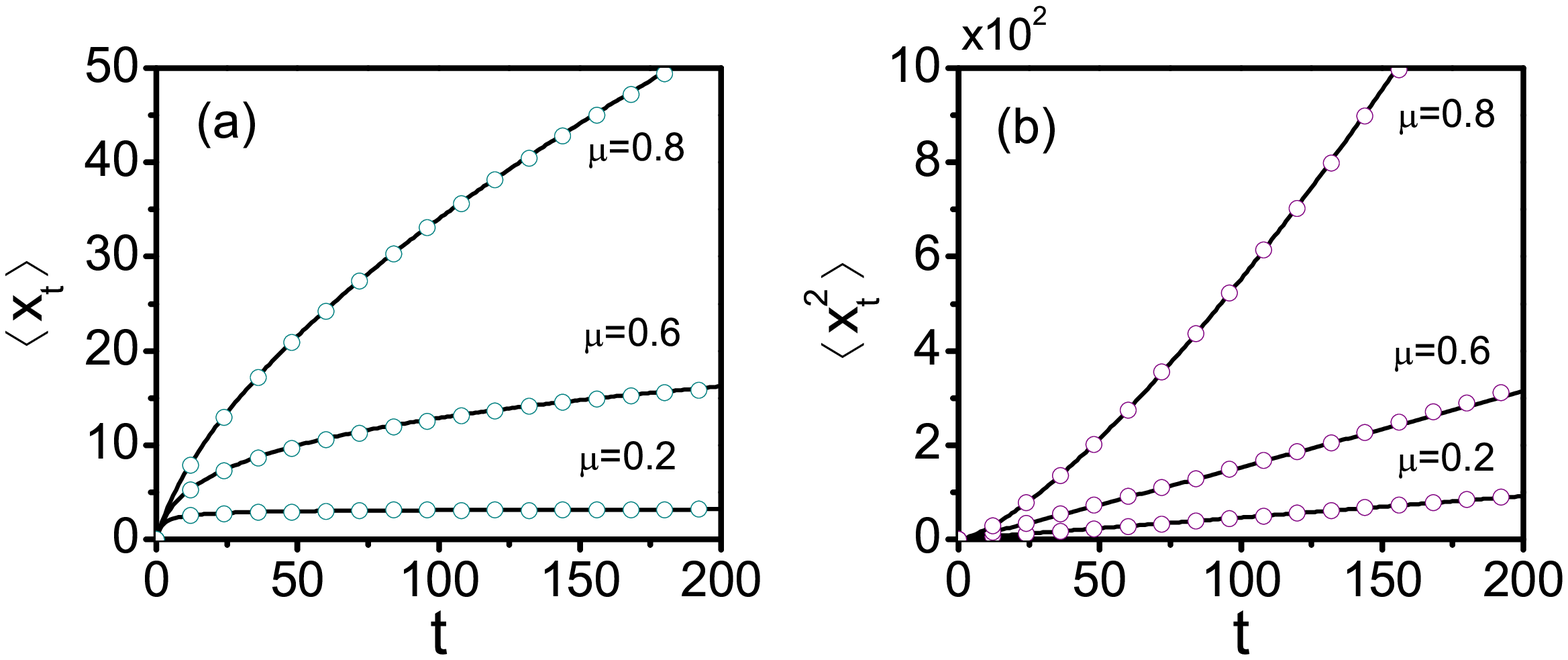}
\caption{Ensemble moments of the random walker. (a) First moment $\langle
x_{t}\rangle .$ (b) Second moment $\langle x_{t}^{2}\rangle .$ The full
lines correspond to the exact expressions (\protect\ref{First}) and (\protect
\ref{Second}) respectively. In (a) we take $q_{+}=1,$ $q_{-}=0,$ while in
(b) $q_{+}=q_{-}=1/2.$ In all cases $\protect\lambda =2.$ The value of $%
\protect\mu $\ is indicated in each curve. Numerical results (circles) were
obtained from an average over $5\times 10^{3}$ realizations. }
\end{figure}
%figura%figura%figura%figura%figurav%figura%figura%figura%figura%figura%figura%figura%figura%figura%figurav%figura%figura%figura%figura%figura
%figura%figura%figura%figura%figurav%figura%figura%figura%figura%figura%figura%figura%figura%figura%figurav%figura%figura%figura%figura%figura

\subsection{Second moment}

For the \textit{second moment,} we get the recursive relation%
\begin{equation}
\langle x_{t+1}^{2}\rangle =\langle x_{t}^{2}\rangle \left[ 1+\frac{2\alpha 
}{t+\lambda }\right] +1+2\delta q\frac{\lambda }{t+\lambda }\langle
x_{t}\rangle ,
\end{equation}%
whose solution is given by%
\begin{equation}
\langle x_{t}^{2}\rangle =\frac{1}{2\alpha -1}\Big{[}\frac{\Gamma (\lambda
+1)}{\Gamma (2\alpha +\lambda )}\frac{\Gamma (2\alpha +\lambda +t)}{\Gamma
(\lambda +t)}-(n+\lambda )\Big{]}+\varphi (t)  \label{Second}
\end{equation}%
where $2\alpha -1=4\mu -3.$ The bracket term gives the solution when $\delta
q=0,$ while $\varphi (t)$ takes into account the contributions proportional
to $\delta q\neq 0,$%
\begin{equation}
\varphi (t)\equiv \delta q^{2}\lambda ^{2}\Big{\{}1+\frac{\Gamma (\lambda )%
\big{[}\frac{\Gamma (\lambda +2\alpha +t)}{\Gamma (\lambda +2\alpha )}-\frac{%
2\Gamma (\lambda +\alpha +t)}{\Gamma (\lambda +\alpha )}\big{]}}{\Gamma
(t+\lambda )}\Big{\}}.  \label{phi}
\end{equation}

In the long time limit, $t\gg \lambda ,$ by using the approximation $\Gamma
(z+v)/\Gamma (z)\simeq z^{v}$ valid for $z\rightarrow \infty ,$ in the case $%
\delta q=0$ we get the asymptotic behaviors%
\begin{equation}
\langle x_{t}^{2}\rangle \approx \left\{ 
\begin{array}{cccc}
\frac{1}{4\mu -3}\frac{\Gamma (\lambda +1)}{\Gamma (4\mu -2+\lambda )}%
t^{4\mu -2} &  &  & \mu >3/4, \\ 
&  &  &  \\ 
t\ln (t) &  &  & \mu =3/4, \\ 
&  &  &  \\ 
\frac{t}{3-4\mu } &  &  & \mu <3/4.%
\end{array}%
\right.  \label{SegundoAsymp}
\end{equation}%
By introducing these behaviors in Eq. (\ref{Delta2Average}) it is possible
to recovers analytically the fitting for $\beta _{2},$ Eq. (\ref{fiteo}). In
fact, corrections proportional to $\langle x_{t}\rangle $ and $\varphi (t),$
Eqs. (\ref{xmedio}) and (\ref{phi}) respectively, gives higher order
(inverse) power-law corrections that can be disregarded in the asymptotic
regime.

We notice that $\langle x_{t}^{2}\rangle $ [Eq. (\ref{SegundoAsymp})]
develops a transition between a\textit{\ normal diffusive behavior} $(\mu
<3/4)$ to a \textit{superdiffusive} one $(\mu >3/4).$ These features are
related to the exponent $\beta _{2}$ of the power-law decay of $\langle
\delta \mathcal{T}_{t}^{2}\rangle ,$ Eqs. (\ref{BetasDefinition}) and (\ref%
{Delta2Average}). In Fig.~4(b) we plot $\langle x_{t}^{2}\rangle $ for
different values of $\mu .$ Numerical simulations confirm the theoretical
predictions. On the other hand, at $\mu =1,$ a \textit{ballistic behavior}
is obtained asymptotically, $\langle x_{t}^{2}\rangle =t(t+\lambda
)/(1+\lambda ),$ result derived in Ref. \cite{urna}.

\subsection{Correlation}

The correlation of the random walker is defined as%
\begin{equation}
C_{t,\tau }^{x}\equiv \langle x_{t}x_{t+\tau }\rangle ,
\end{equation}%
with initial condition $C_{t,0}^{x}=\langle x_{t}^{2}\rangle .$ From the
double characteristic function [Appendix \ref{QFourier}], it is possible to
obtain the recursive relation%
\begin{equation}
C_{t,\tau +1}^{x}=C_{t,\tau }^{x}\left[ 1+\frac{\alpha }{t+\tau +\lambda }%
\right] +\frac{\lambda \delta q}{t+\tau +\lambda }\langle x_{t}\rangle .
\label{RecursivaCorrelationCorre}
\end{equation}%
Its solution is%
\begin{equation}
C_{t,\tau }^{x}=\Big{[}\langle x_{t}^{2}\rangle +\frac{\lambda \delta q}{%
\alpha }\langle x_{t}\rangle \Big{]}\Phi (t,\tau )-\frac{\lambda \delta q}{%
\alpha }\langle x_{t}\rangle ,  \label{Corre}
\end{equation}%
where $\langle x_{t}\rangle $\ and $\langle x_{t}^{2}\rangle $\ follow from
Eqs. (\ref{First}) and (\ref{Second}) respectively. The auxiliary function $%
\Phi (t,\tau )$ reads%
\begin{equation}
\Phi (t,\tau )\equiv \frac{\Gamma (\lambda +t)}{\Gamma (\alpha +\lambda +t)}%
\frac{\Gamma (\alpha +\lambda +t+\tau )}{\Gamma (\lambda +t+\tau )}.
\end{equation}

For $\mu =1/2,$ that is $\alpha =0,$ the solution of the recursive relation (%
\ref{RecursivaCorrelationCorre}) reads $C_{t,\tau }^{x}=\langle
x_{t}^{2}\rangle +\langle x_{t}\rangle \lambda \delta q[\psi (\lambda
+t+\tau )-\psi (t+\tau )],$ where the digamma function is defined as $\psi
(z)=(d/dz)\log \Gamma (z).$ In the limit $\mu \rightarrow 1,$ $C_{t,\tau
}^{x}$ is given by Eq. (\ref{Corre}) with $\Phi (t,\tau )=[1+\tau
/(t+\lambda )].$

In the asymptotic regime, by using the approximation $\Gamma (z+v)/\Gamma
(z)\simeq z^{v},$ valid for $z\rightarrow \infty ,$ it follows $\Phi (t,\tau
)\simeq \lbrack 1+\tau /(t+\lambda )]^{\alpha },$ which leads to%
\begin{equation}
C_{t,\tau }^{x}\simeq \Big{[}\langle x_{t}^{2}\rangle +\frac{\lambda \delta q%
}{\alpha }\langle x_{t}\rangle \Big{]}\Big{(}1+\frac{\tau }{t}\Big{)}%
^{\alpha }-\frac{\lambda \delta q}{\alpha }\langle x_{t}\rangle .
\label{CorreAsymp}
\end{equation}%
Thus, in the asymptotic regime the correlation depends on the quotient $%
(\tau /t),$ showing the strong non-stationarity property of the diffusion
process. A similar result is also valid for Levy walks \cite{einsteinLevy}.
On the other hand, for $\mu \gtrless 1/2$ (equivalently $\alpha \gtrless 0)$ 
$C_{t,\tau }^{x}$ increases (decrees) with $\tau .$ This result is
consistent with the correlation-anticorrelation mechanism introduced by $\mu
.$ In order to check these results, in Fig. (5) we plot the normalized
correlation $\langle x_{t}x_{t+\tau }\rangle /\langle x_{t}^{2}\rangle $ as
a function of the interval $\tau .$ Numerical simulations and analytical
results are indistinguishable in the scale of the plots.%
%figura1%figura%figura%figura%figurav%figura%figura%figura%figura%figura%figura%figura%figura%figura%figurav%figura%figura%figura%figura%figura
%figura%figura%figura%figura%figurav%figura%figura%figura%figura%figura%figura%figura%figura%figura%figurav%figura%figura%figura%figura%figura
\begin{figure}[tbp]
\includegraphics[bb=50 805 490 1140,angle=0,width=6.5cm]{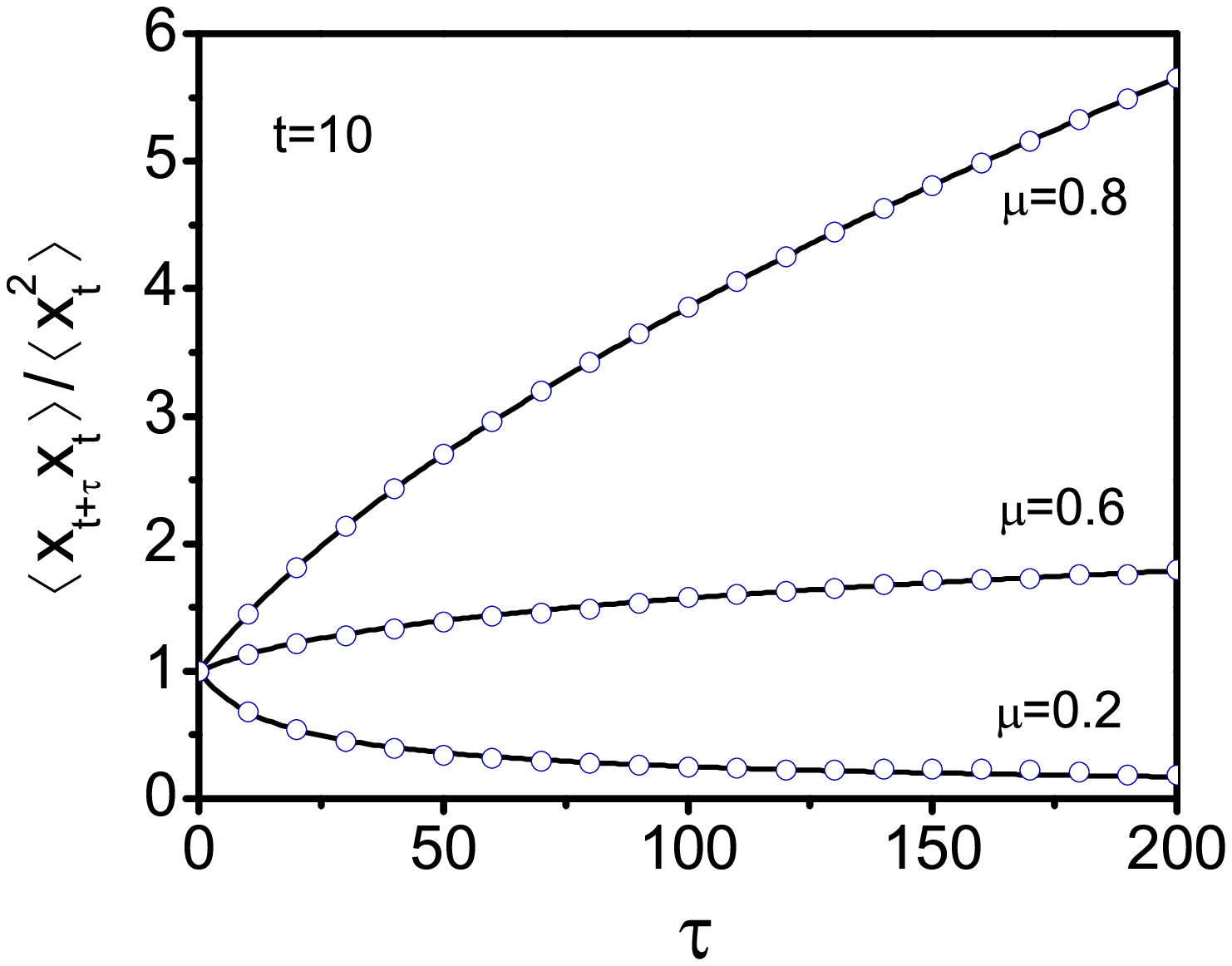}
\caption{Normalized correlation $\langle x_{t}x_{t+\protect\tau }\rangle
/\langle x_{t}^{2}\rangle $ as a function of the difference time $\protect%
\tau .$ The full lines correspond to the exact result Eq. (\protect\ref%
{Corre}). In all cases we take $t=10,$ $q_{+}=q_{-}=1/2,$ and $\protect%
\lambda =2.$ The value of $\protect\mu $\ is indicated in each curve.
Numerical results (circles) were obtained from an average over $2\times
10^{4}$ realizations.}
\end{figure}
%figura%figura%figura%figura%figurav%figura%figura%figura%figura%figura%figura%figura%figura%figura%figurav%figura%figura%figura%figura%figura
%figura%figura%figura%figura%figurav%figura%figura%figura%figura%figura%figura%figura%figura%figura%figurav%figura%figura%figura%figura%figura

\subsection{Joint-probability evolution}

By Fourier inversion, $k_{1}\rightarrow y,$ $k_{2}\rightarrow x,$ the double
characteristic function (\ref{doble}) also allows us to obtain the joint
probability $P(y,t;x,\tau )$ of observing the walker at position $y$ at time 
$t$ and at position $x$ at time $t+\tau .$ From Eq. (\ref{DoubleRecursive})
we get%
\begin{eqnarray}
P(y,t;x,\tau +1) &=&W_{t,\tau }^{+}(x-1)P(y,t;x-1,\tau )  \notag \\
&&+W_{t,\tau }^{-}(x+1)P(y,t;x+1,\tau ),\ \ \ \ \ \ \ 
\end{eqnarray}%
where the transition probabilities are%
\begin{equation}
W_{t,\tau }^{\pm }(x)=\frac{1}{2}\left[ 1\pm \frac{1}{t+\tau +\lambda }%
(\alpha x+\lambda \delta q)\right] .  \label{Rates}
\end{equation}%
Hence, the dynamics as a function of the interval $\tau $ develops aging 
\cite{aging,xxx,age}, that is, here the transition probabilities $W_{t,\tau
}^{\pm }(x)$ depend on the starting time $t$ with a power-law dependence.
This property is closely related with the asymptotic power-law behavior of
the walker correlation, Eq. (\ref{CorreAsymp}). These features are absent in
the memoryless limit, $\lim_{\lambda \rightarrow \infty }W_{t,\tau }^{\pm
}(x)=q_{\pm }.$

In a continuous limit, where both the jump length $\delta x$ and the time
interval $\delta t$ between consecutive transitions\ become small, the
previous master equation leads to the Focker-Planck equation%
\begin{eqnarray*}
\frac{\partial }{\partial \tau }P(y,t;x,\tau ) &=&D\frac{\partial ^{2}}{%
\partial ^{2}x}P(y,t;x,\tau ) \\
&&-\frac{\alpha }{t+\tau +t_{\lambda }}\frac{\partial }{\partial x}%
[xP(y,t;x,\tau )] \\
&&-\frac{t_{\lambda }}{t+\tau +t_{\lambda }}V\frac{\partial }{\partial x}%
P(y,t;x,\tau ),
\end{eqnarray*}%
where the parameters are $D\equiv (1/2)\delta x^{2}/\delta t,$ $V\equiv
(q_{+}-q_{-})(\delta x/\delta t),$ and $t_{\lambda }\equiv \lambda \delta t.$
Interestingly, this equation has the form of a diffusion process in a time-
and age- (power-law) dependent inverted parabolic potential superimposed
with a time-dependent linear drift. Notice that around $\mu =1/2$ the
potential is inverted, property related to the correlation-anticorrelation
mechanism introduced by the parameter $\mu .$

\section{Time-averaged observables}

Here we study the ergodic properties of the walker defined by Eq. (\ref%
{Transition}). Given the discrete nature of the dynamics, the definition of
the time-averaged moments is given by%
\begin{equation}
\delta _{\kappa }(t,\Delta )=\frac{\sum_{t^{\prime }=0}^{t-\Delta
}[x(t^{\prime }+\Delta )-x(t^{\prime })]^{\kappa }}{t-\Delta }.
\label{TimeAveraged}
\end{equation}%
Here, we have used the traslational invariance of Eq. (\ref{Definition}),
which allow to write the definition in terms of $x(t)$ [Eq. (\ref{position}%
)]. The definitions of ergodicity and ergodicity breaking are those quoted
in the Introduction.

\subsection{Infinite-time trajectories}

The walker ensemble properties are mainly determinate by the decay behavior
of the transition probabilities. In contrast, for infinite-time
trajectories, $\lim_{t\rightarrow \infty }\delta _{\kappa }(t,\Delta ),$ the
time-averaged moments are settled by the asymptotic behavior of the
transition probabilities. In fact, taking higher times $t$ in Eq. (\ref%
{TimeAveraged}), the relevant walker transitions are those governed by the
asymptotic value $\mathcal{T}_{\infty }^{\pm }=1/2$ [Eq. (\ref{UnMedio})].
Consequently, we expect that along a single trajectory the time averaged
moments of $x_{t}$ converge to those of an undriven normal diffusion
process. Hence, it follows%
\begin{equation}
\lim_{t\rightarrow \infty }\delta _{1}(t,\Delta )=0,\ \ \ \ \ \ \ \ 0\leq
\mu <1,  \label{Delta1Asymp}
\end{equation}%
while the time-averaged mean square displacement reads%
\begin{equation}
\lim_{t\rightarrow \infty }\delta _{2}(t,\Delta )=\Delta ,\ \ \ \ \ \ \ \
0\leq \mu <1.  \label{Delta2Asymp}
\end{equation}%
For normal diffusion, these expressions follows straightforwardly from the
definition (\ref{TimeAveraged}) and by considering independent random walker
increments.

The previous results imply that the process is not ergodic. In fact, $%
\langle x_{\Delta }\rangle \neq \lim_{t\rightarrow \infty }\delta
_{1}(t,\Delta ),$ and $\langle x_{\Delta }^{2}\rangle \neq
\lim_{t\rightarrow \infty }\delta _{2}(t,\Delta ).$ These inequalities
remain valid even for $\Delta \gg 1$ [see Eqs. (\ref{xmedio}) and (\ref%
{SegundoAsymp}) respectively] and are valid for any value of $\lambda ,$ $%
\mu ,$ and $q_{\pm }.$

Interestingly, while the bias induced by $\delta q=q_{+}-q_{-}$ drives the
ensemble behavior [see Eq. (\ref{xmedio})], the time-averaged response $%
[\lim_{t\rightarrow \infty }\delta _{1}(t,\Delta )]$ vanishes (dye out)
asymptotically [Eq. (\ref{Delta1Asymp})]. In consequence, it is not possible
to ask about an Einstein fluctuation dissipation relation formulated with
time-averaged observables (infinite-time trajectories). This unusual
property relies on both the power-law decay of the transitions probabilities
and their stationary on-half values.

While for the second moment time and ensemble averages are always different,
when $\mu <3/4$ [diffusive-like regime, Eq. (\ref{SegundoAsymp})] they
differ only in terms of a constant, $\lim_{t\rightarrow \infty }\delta
_{2}(t,\Delta )/(3-4\mu )=\Delta /(3-4\mu )\simeq \langle x_{\Delta
}^{2}\rangle .$ In contrast, for Levy walks this ultraweak ergodicity
breaking is valid in the superdiffusive regime.

In order to check these results, in Fig.~6 we plot a set of realizations
corresponding to $\delta _{2}(t,\Delta )$ for different values of $\mu .$
Around the origin all realizations approach the limit defined by Eq. (\ref%
{Delta2Asymp}). We checked that by increasing the measurement time $t,$ the
departure with respect to the linear behavior consistently occurs at larger
delay times $\Delta .$%
%figura1%figura%figura%figura%figurav%figura%figura%figura%figura%figura%figura%figura%figura%figura%figurav%figura%figura%figura%figura%figura
%figura%figura%figura%figura%figurav%figura%figura%figura%figura%figura%figura%figura%figura%figura%figurav%figura%figura%figura%figura%figura
\begin{figure}[tbp]
\includegraphics[bb=45 590 722 1133,angle=0,width=8.5cm]{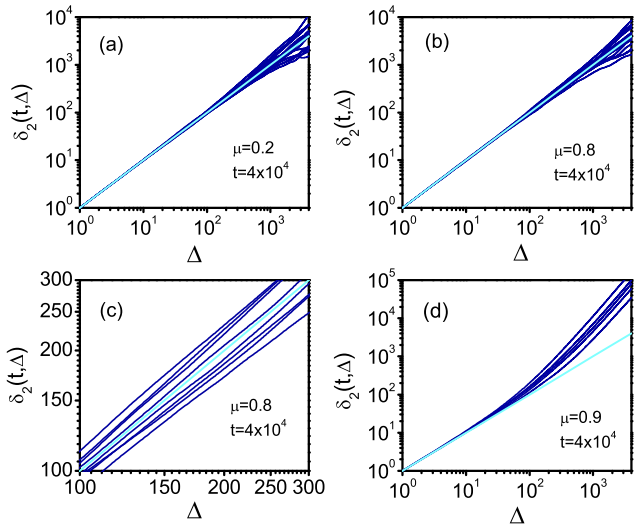}
\caption{Dependence with the lag time $\Delta $\ of the time-averaged mean
square displacement $\protect\delta _{2}(t,\Delta )$ obtained for different
walker trajectories [see Fig. (1)]. The gray lines (light blue lines)
correspond to the infinite trajectory limit, $\lim_{t\rightarrow \infty }%
\protect\delta (t,\Delta )=\Delta .$ In (a) (25 trajectories) we take $%
\protect\mu =0.2.$ In (b) (25 trajectories) $\protect\mu =0.8.$ In (c) a few
of the previous trajectories are shown in the time scale posterior to the
linear regime. In (d) (5 trajectories) $\protect\mu =0.9.$ In all cases, we
take $\protect\lambda =2,$ $q_{+}=q_{-}=1/2,$ and $t=4\times 10^{4}.$}
\end{figure}
%figura%figura%figura%figura%figurav%figura%figura%figura%figura%figura%figura%figura%figura%figura%figurav%figura%figura%figura%figura%figura
%figura%figura%figura%figura%figurav%figura%figura%figura%figura%figura%figura%figura%figura%figura%figurav%figura%figura%figura%figura%figura

\subsection{Randomness in finite-time trajectories}

The previous results are valid for any (finite) value of $\lambda $ and $%
0\leq \mu <1.$ When $\mu =1,$ the model reduces to the urn-like dynamics of
Ref. \cite{urna}. Thus,%
\begin{equation}
\lim_{t\rightarrow \infty }\delta _{1}(t,\Delta )=(f_{+}-f_{-})\Delta ,
\label{Delta1Random}
\end{equation}%
while the second time-averaged moment reads%
\begin{equation}
\lim_{t\rightarrow \infty }\delta _{2}(t,\Delta )=(f_{+}-f_{-})^{2}\Delta
^{2}+[1-(f_{+}-f_{-})^{2}]\Delta ,  \label{Delta2Random}
\end{equation}%
where $f_{\pm }$ are Beta random variables, with $f_{+}+f_{-}=1$ (see Eqs.
(32) and (33) in Ref. \cite{urna} where these results were derived). The
transition between these scaling and those defined by Eqs. (\ref{Delta1Asymp}%
) and (\ref{Delta2Asymp}) can be described by analyzing the behavior of the
time-averaged moments obtained with finite-time trajectories.

In the plots of Fig.~6 we observe that, even when $\Delta \ll t,$ beyond the
linear regime the scaling of $\delta _{2}(t,\Delta )$ can be subdiffusive or
superdiffusive. Furthermore, the amplitude of the scaling can also be random
[see Fig.~6(c)]. These properties also arise in Levy walks \cite{godec}.
Here, these features are present for all values of $\mu .$ Hence, we
associate these effects to the random behavior of the transition
probabilities (see Fig.~1 and 2). In fact, independently of the values of
the memory parameters $\mu $ and $\lambda ,$\ they decay to their stationary
values following a power-law behavior with parameters that are intrinsically
random [see Eqs. (\ref{DeltaTran}) and (\ref{BetasDefinition})]. For $\mu
\approx 1,$ all realizations becomes superdiffusive [see Fig.~6(d), $\mu
=0.9],$ a consistent property necessary\ for approaching the scaling defined
by Eq.~(\ref{Delta2Random}).

\subsection{Ensemble average of finite-time trajectories}

Now we study how the finiteness of single trajectories affects the
corresponding average over an ensemble of trajectories. From Eq. (\ref%
{TimeAveraged}) the first moment reads%
\begin{equation}
\langle \delta _{1}(t,\Delta )\rangle =\frac{1}{t-\Delta }\sum_{t^{\prime
}=0}^{t-\Delta }\langle x_{t^{\prime }+\Delta }\rangle -\langle x_{t^{\prime
}}\rangle ,  \label{Delta1Discreto}
\end{equation}%
while the second one can be written as%
\begin{equation}
\langle \delta _{2}(t,\Delta )\rangle =\frac{1}{t-\Delta }\sum_{t^{\prime
}=0}^{t-\Delta }\langle x_{t^{\prime }+\Delta }^{2}\rangle +\langle
x_{t^{\prime }}^{2}\rangle -2\langle x_{t^{\prime }+\Delta }x_{t^{\prime
}}\rangle .  \label{Delta2Discreto}
\end{equation}

Given the exact analytical expressions for $\langle x_{t}\rangle $ [Eq. (\ref%
{First})], $\langle x_{t}^{2}\rangle $ [Eq. (\ref{Second})], and the
correlation $\langle x_{t^{\prime }+\Delta }x_{t^{\prime }}\rangle $ [Eq. (%
\ref{Corre})], we can also evaluate these objects in an exact way.
Nevertheless, they cannot be expressed in terms of general simple
expressions. Only for special values one get simpler ones. For example, for $%
\mu =1$ Eq. (\ref{Delta1Discreto}) becomes%
\begin{equation}
\langle \delta _{1}(t,\Delta )\rangle =\delta q\Delta \Big{[}1+\frac{1}{%
t-\Delta }\Big{]},\ \ \ \ \ \ \ \ \mu =1.
\end{equation}%
Taking $\delta q=q_{+}-q_{-}=0,$ the mean square displacement [Eq. (\ref%
{Delta2Discreto})] reads 
\begin{equation}
\langle \delta _{2}(t,\Delta )\rangle =\frac{\Delta (\Delta +\lambda )}{%
1+\lambda }\Big{[}1+\frac{1}{t-\Delta }\Big{]},\ \ \ \ \ \ \ \mu =1.
\label{Delta2ExactoMu1}
\end{equation}%
In the limit $t\rightarrow \infty ,$ these expressions correspond to the
average over realizations of Eqs. (\ref{Delta1Random}) and (\ref%
{Delta2Random}) (see Ref. \cite{urna}). On the other hand, the finite-time
effects are given by the contributions proportional to $1/(t-\Delta ).$

For $\mu <1,$ given that not simple analytical expression can be obtained,
in Appendix \ref{SumaAprox} we introduce a set of approximations that allow
to obtaining the asymptotic behavior $(t\gg \Delta )$ of the exact
expressions Eqs. (\ref{Delta1Discreto}) and (\ref{Delta2Discreto}). We get%
\begin{equation}
\langle \delta _{1}(t,\Delta )\rangle \sim \delta q\,c_{0}\Big{\{}\frac{2\mu
\Delta }{t^{2(1-\mu )}}-\frac{1}{t}[(\Delta +\lambda )^{2\mu }-\lambda
^{2\mu }]\Big{\}},  \label{AjusteDelta1}
\end{equation}%
where $c_{0}=\Gamma (\lambda +1)/[2\mu \alpha \Gamma (\alpha +\lambda )].$
Consistently with Eq. (\ref{Delta1Asymp}), $\langle \delta _{1}(t,\Delta
)\rangle $ vanishes when $t\rightarrow \infty .$ This regime is approached
following a power-law behavior. In fact, for $\mu <1/2,\ \langle \delta
_{1}(t,\Delta )\rangle \sim \Delta ^{2\mu }/t,$ while for $\mu >1/2,\
\langle \delta _{1}(t,\Delta )\rangle \sim \Delta /t^{2(1-\mu )}.$

Taking $\delta q=q_{+}-q_{-}=0,$ for the mean squared displacement we obtain%
\begin{eqnarray}
\langle \delta _{2}(t,\Delta )\rangle &\sim &\Delta +\frac{\Delta ^{2}}{t}%
\Big{[}a+b\ln \Big{(}\frac{\Delta }{t}\Big{)}\Big{]}  \notag \\
&&+c\frac{\Delta ^{2}}{t^{4(1-\mu )}}+d\frac{\Delta ^{4\mu -1}}{t},
\label{AjusteDelta2}
\end{eqnarray}%
where $a,$ $b,$ $c,$ and $d$ are constants given in Appendix \ref{SumaAprox}%
. The factor proportional to $d$ only contributes for $\mu >1/2.$ 
%figura1%figura%figura%figura%figurav%figura%figura%figura%figura%figura%figura%figura%figura%figura%figurav%figura%figura%figura%figura%figura
%figura%figura%figura%figura%figurav%figura%figura%figura%figura%figura%figura%figura%figura%figura%figurav%figura%figura%figura%figura%figura
\begin{figure}[tbp]
\includegraphics[bb=45 866 730 1145,angle=0,width=8.5cm]{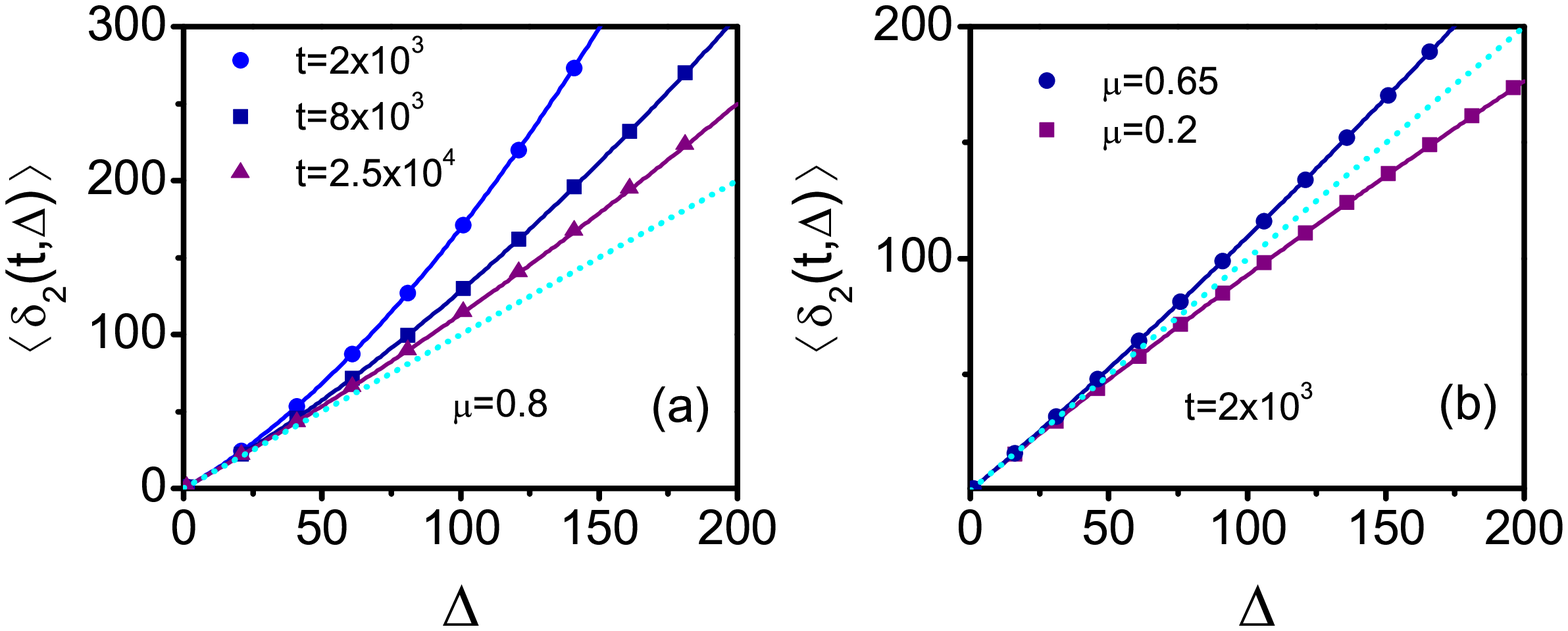}
\caption{Ensemble average $\langle \protect\delta _{2}(t,\Delta )\rangle $
as a function of $\Delta .$ In (a) we take $\protect\mu =0.8,$ an three
different time, $t=2\times 10^{3}$ (circles), $t=8\times 10^{3}$ (squares),
and $t=25\times 10^{3}$ (triangles). In (b) we take $t=2\times 10^{3}$ and $%
\protect\mu =0.65$ (circles), $\protect\mu =0.2$ (squares). The full lines
correspond to the fitting (\protect\ref{AjusteDelta2}) valid for large
times. The dotted lines give the infinite trajectory limit, $%
\lim_{t\rightarrow \infty }\langle \protect\delta _{2}(t,\Delta )\rangle
=\Delta .$ The numerical results (circles, squares, triangles) \ were
obtained from an average over $2\times 10^{3}$ realizations.}
\end{figure}
%figura%figura%figura%figura%figurav%figura%figura%figura%figura%figura%figura%figura%figura%figura%figurav%figura%figura%figura%figura%figura
%figura%figura%figura%figura%figurav%figura%figura%figura%figura%figura%figura%figura%figura%figura%figurav%figura%figura%figura%figura%figura

In Eq. (\ref{AjusteDelta2}), the deviations with respect to the linear
behavior (\ref{Delta2Asymp}) change around $\mu =3/4.$ For $\mu <3/4,$ the
dominant terms are those proportional to $a$ and $b,$ while for $\mu >3/4$
are those proportional to $c$ and $d.$ In fact, the quadratic contribution $%
\Delta ^{2}$ dominates at $\mu =1,$ which approximate the exact behavior (%
\ref{Delta2ExactoMu1}).

In contrast to Levy walks (see Eq. (18) In Ref. \cite{einsteinLevy}), by
comparing the asymptotic behaviors of $\langle \delta _{1}(t,\Delta )\rangle 
$ [Eq. (\ref{AjusteDelta1})] and $\langle \delta _{2}(t,\Delta )\rangle $
[Eq. (\ref{AjusteDelta2})] we conclude that, for finite-time trajectories,
it is not possible to establishing a simple relation between both objects
(Einstein-like relation).

In order to check the previous results, in Fig.~7(a) we plot $\langle \delta
_{2}(t,\Delta )\rangle $ for different times and the same $\mu .$ An
increasing convergence to the lineal regime, $\langle \delta _{2}(t,\Delta
)\rangle \simeq \Delta ,$ is observed for increasing $t.$ In Fig.~7(b) we
plot $\langle \delta _{2}(t,\Delta )\rangle $ for different values of $\mu .$
For $\mu <1/2$ it is a concave function of $\Delta $ while convex for $\mu
>1/2.$ In all plots, the numerical results, the exact result (\ref%
{Delta2Discreto}) and the approximation (\ref{AjusteDelta2}) are
indistinguishable in the scale of the graphs.

\section{Summary and Conclusions}

The studied model consists of a diffusive walker whose successive jumps
depend on the whole previous history of transitions [Eq. (\ref{Transition}%
)]. The second moment develops a diffusive-superdiffusive transition. This
memory-induced property can be directly related to a transition in the
power-law decay behavior of the transition probabilities to their stationary
values (see Fig. 1 to 3), which in fact develop a similar transition for the
same parameter values [see Eqs. (\ref{fiteo}) and (\ref{SegundoAsymp})]. The
ensemble behavior is non-stationary and develops aging, that is, the
transition probabilities governing the walker ensemble depend on the initial
time [Eq. (\ref{Rates})]. The random drift induced by the difference between
the transitions probabilities lead to trajectories where the walker may
persist during an entire realization with the same velocity. Nevertheless,
the time intervals where this happen are characterized by a finite average
(Appendix A).

Given that the transition probabilities asymptotically converge to one half
[Eq. (\ref{UnMedio})], time-averaged moments performed with infinite-time
trajectories become equivalent to that of an unbiased normal random walk
[Eqs. (\ref{Delta1Asymp}) and (\ref{Delta2Asymp})]. Hence, the process is
non-ergodic. The vanishing of the first time-averaged moment implies that
the dynamics is (asymptotically) insensitive to the bias introduced by the
characteristic parameters. On the other hand, in the diffusive regime an
ultraweak ergodicity breaking phenomenon occurs, that is, for the second
moment ensemble and time-averages only differ by a constant.

For finite-time trajectories, the time-averaged moments develop a randomness
that appears in both the scaling exponents and their amplitudes (Fig.~6).
This effect is induced by the intrinsic randomness of the power-law decay of
the transition probabilities (Fig.~1). Departure between ensemble averages
performed with finite-time trajectories (Fig.~7) and the corresponding
infinite-time limit are also governed by power-law behaviors [Eqs. (\ref%
{AjusteDelta1}) and (\ref{AjusteDelta2})]. None simple relation can be
established between the mean asymptotic behaviors of the first two
time-averaged moments (driven and undriven cases).

The studied model recovers many features that also arise in Levy walks.
While their time-averaged properties are not equivalent, the present results
demonstrate that many properties of anomalous diffusive processes can also
be recovered with simple globally correlated dynamics.

\section*{Acknowledgments}

This work was supported by Consejo Nacional de Investigaciones Cient\'{\i}%
ficas y T\'{e}cnicas (CONICET), Argentina.

\appendix

\section{\label{TIMES}Probability of sojourn times}

Here, we obtain the probability of the sojourn times, that is, the time
intervals during which the walker moves in the same direction. Equivalently,
they correspond to the time during which the \textquotedblleft
velocity\textquotedblright\ is the same. Given that the walker at time $t$
performed $t_{\pm }$ right-left transitions, the probability of performing $%
k $ successive jumps to the \textit{right}, from the definition (\ref%
{Transition}), is given by%
\begin{eqnarray}
P_{t}(k) &=&\prod_{i=0}^{k-1}\frac{\lambda q_{+}+\mu (t_{+}+i)+(1-\mu )t_{-}%
}{t+i+\lambda }  \notag \\
&&\times \frac{\lambda q_{-}+\mu t_{-}+(1-\mu )(t_{+}+k)}{t+k+\lambda }.
\end{eqnarray}%
The last term takes into account the beginning of a sojourn time with
transitions in the opposite direction. By using the property of the Gamma
function: $\Gamma (n+z)/\Gamma (z)=z(1+z)(2+z)\cdots (n-1+z),$ the previous
equation becomes%
\begin{equation}
P_{t}(k)=\frac{\Gamma (t+\lambda )}{\Gamma (t+\lambda +k+1)}[\tau
_{-}+(1-\mu )k]\mu ^{k}\frac{\Gamma (k+\tau _{+}/\mu )}{\Gamma (\tau
_{+}/\mu )},  \label{Peka}
\end{equation}%
where for shortening the expression we introduced the parameters%
\begin{equation}
\tau _{\pm }\equiv \lambda q_{\pm }+\mu t_{\pm }+(1-\mu )t_{\mp }=\lambda
q_{\pm }+\frac{1}{2}(t\pm \alpha x_{t}),
\end{equation}%
$\alpha =(2\mu -1).$ The last equality straightforwardly follows\ from the
relation (\ref{tiempos}). In this way, $P_{t}(k)$ depends on which time and
which position the sojourn interval begins. The structure of this dependence
is simpler in the asymptotic regime $t\gg \lambda .$ By using the Gamma
function property $\Gamma (z+v)/\Gamma (z)\simeq z^{v}$\ valid for $%
z\rightarrow \infty ,$ Eq. (\ref{Peka}) becomes%
\begin{equation}
P_{t}(k)\simeq \frac{1}{(t+\lambda )^{k+1}}[\tau _{-}+(1-\mu )k]\tau
_{+}^{k}.
\end{equation}%
Approximating $\tau _{\pm }\simeq tw_{\pm }$ for $t\gg \lambda ,$ where%
\begin{equation}
w_{\pm }\equiv \frac{1}{2}[1\pm \alpha \frac{x_{t}}{t}],
\end{equation}%
$(w_{+}+w_{-}=1)$ we obtain the final expression%
\begin{equation}
P_{t}(k)\simeq w_{+}^{k}w_{-}.  \label{bino}
\end{equation}%
The corrections to this expression are of order $(1/t).$ On the other hand,
we notice that $w_{\pm }$ are the asymptotic transition probabilities
defined in Eq. (\ref{TranAsimp}).

For $\mu \gtrless 1/2,$ for increasing (decreasing) $x_{t}$ the probability $%
P_{t}(k)$ increase (decrease). That is, if the particle attains larger
(smaller) values of $x_{t}$ the possibility of larger sojourn times in the
same direction increase (decrease). This dependence of $P_{t}(k)$ with $%
x_{t} $ is confirmed by its values in the boundary $x_{t}=\pm t,$ and $%
x_{t}=0,$%
\begin{equation}
P_{t}(k)\simeq \left\{ 
\begin{array}{c}
(1-\mu )\mu ^{k},\ \ \ \ \ \ \ \ \ \ \ \ x_{t}=+t, \\ 
\\ 
(\frac{1}{2})^{k+1},\ \ \ \ \ \ \ \ \ \ \ \ \ \ x_{t}=0, \\ 
\\ 
\mu (1-\mu )^{k},\ \ \ \ \ \ \ \ \ \ \ \ x_{t}=-t,%
\end{array}%
\right.
\end{equation}%
which in turn also clarify the role of the parameter $\mu $ in the walker
realizations. Interestingly, in spite of the previous feature the average
sojourn time is finite $\langle k\rangle \equiv \sum_{k=0}^{\infty
}kP_{t}(k)\simeq w_{+}/(1-w_{+}),$ as well as the second moment, $\langle
k^{2}\rangle \equiv \sum_{k=0}^{\infty }k^{2}P_{t}(k)\simeq
w_{+}(1+w_{+})/(1-w_{+})^{2}.$ Straightforwardly, the same results apply for
the probability of sojourn times in the opposite direction.

\section{\label{QFourier}Double characteristic function}

In this Appendix we obtain an exact recursive relation for the double
characteristic function%
\begin{equation}
Q(k_{1},t;k_{2},\tau )\equiv \langle \exp [i(k_{1}x_{t}+k_{2}x_{t+\tau
})]\rangle .  \label{Characteristic}
\end{equation}%
It is obtained as follows. At time $\tau +1,$ it can be written as%
\begin{eqnarray}
Q(k_{1},t;k_{2},\tau +1) &=&\Big{\langle}\exp [i(k_{1}x_{t}+k_{2}x_{t+\tau
})] \\
&&\times \sum_{\sigma =\pm 1}e^{ik_{2}\sigma }\mathcal{T}(\sigma _{1},\cdots
\sigma _{t+\tau }|\sigma )\Big{\rangle}.  \notag
\end{eqnarray}%
Here, we have taken into account that the random variable $\sigma _{t+\tau
+1}$ is chosen in agreement with the transition probability $\mathcal{T}%
(\sigma _{1},\cdots \sigma _{t+\tau }|\sigma _{t+\tau +1}).$ Notice that the
ensemble average $\langle \cdots \rangle $ includes all possible random
values of $\{\sigma _{i}\}_{i=1}^{i=t+\tau },$ which in turn define all
possible realizations of $x_{t}$ and $x_{t+\tau }.$ From Eq.~(\ref%
{Transition}), we get%
\begin{eqnarray}
Q_{t,\tau +1} &=&\frac{1}{t+\tau +\lambda }\Big{[}\lambda Q_{t,\tau
}\sum_{\sigma =\pm 1}q_{\sigma }e^{ik_{2}\sigma }  \label{QdobleAveraged} \\
&&+\sum_{\sigma =\pm 1}\Big{\langle}\exp [i(k_{1}x_{t}+k_{2}x_{t+\tau
})]U_{t+\tau }^{\sigma }\Big{\rangle}e^{ik_{2}\sigma }\Big{]},  \notag
\end{eqnarray}%
where the change of notation $Q(k_{1},t;k_{2},\tau )\rightarrow Q_{t,\tau }$
was introduced for shortening the expression. Furthermore, the random
function $U_{t+\tau }^{\sigma }$\ is defined as%
\begin{equation}
U_{t}^{\pm }\equiv \mu t_{\pm }+(1-\mu )t_{\mp },
\end{equation}%
which due to the relation \ $t_{\pm }=(t\pm x_{t})/2$ [Eq. (\ref{tiempos})]
can be rewritten as $U_{t}^{\pm }=(t\pm \alpha x_{t})/2.$ Using that%
\begin{equation}
\frac{\partial Q_{t,\tau }}{\partial k_{2}}=i\langle x_{t+\tau }\exp
[i(k_{1}x_{t}+k_{2}x_{t+\tau })]\rangle ,  \label{derivada}
\end{equation}%
jointly with the normalization condition $q_{+}+q_{-}=1,$ after some
algebra, from Eq. (\ref{QdobleAveraged}) it follows the closed recursive
relation%
\begin{eqnarray}
Q_{t,\tau +1} &=&\cos (k_{2})Q_{t,\tau }+i\lambda \delta q\frac{\sin (k_{2})%
}{(t+\tau +\lambda )}Q_{t,\tau }  \notag \\
&&+\alpha \frac{\sin (k_{2})}{(t+\tau +\lambda )}\frac{\partial Q_{t,\tau }}{%
\partial k_{2}}  \label{DoubleRecursive}
\end{eqnarray}%
where $\alpha =(2\mu -1),$ and$\ \delta q=(q_{+}-q_{-}).$ By differentiation
with respect to $k_{1}$\ and $k_{2}$ the recursive relations presented in
Sec. III\ follows straightforwardly.

\section{\label{SumaAprox}Approximation for the ensemble time-averaged
moments}

Here we show the procedure to obtain the asymptotic behavior of the
time-averaged moments $\langle \delta _{\kappa }(t,\Delta )\rangle .$ Their
exact expressions, Eqs. (\ref{Delta1Discreto}) and (\ref{Delta2Discreto}),
have the following structure%
\begin{equation}
F(t,\Delta )=\frac{1}{t-\Delta }\sum_{t^{\prime }=0}^{t-\Delta }f(t^{\prime
},\Delta ).
\end{equation}%
The goal is to approximate $F(t,\Delta )$ at large time scales, $t\gg \Delta
,$ given that we have an exact expression for $f(t^{\prime },\Delta )$
written in terms of the ensemble moments $\langle x_{t}\rangle ,$ $\langle
x_{t}^{2}\rangle ,$ and the correlation $\langle x_{t^{\prime }+\Delta
}x_{t^{\prime }}\rangle .$

Defining the variable $\varepsilon \equiv \Delta /t,$ the scaled time $\tau
\equiv t^{\prime }/t,$ and $d\tau \equiv 1/t,$ the previous general
expression can be rewritten as%
\begin{equation}
F(t,t\varepsilon )=\frac{1}{1-\varepsilon }\sum_{\tau =0}^{1-\varepsilon
}f(t\tau ,t\varepsilon )d\tau .
\end{equation}%
In this expression $d\tau =1/t\ll 1,$ and consistently $\tau =t^{\prime }/t$
can be considered as a real continuous variable. Therefore, we can
approximate the sum by an integral,%
\begin{equation}
F(t,\Delta )\simeq \frac{1}{1-\varepsilon }\int_{0}^{1-\varepsilon }f(t\tau
,t\varepsilon )d\tau .
\end{equation}%
In addition, given that only the asymptotic regime is of interest, before
performing this integral $f(t\tau ,t\varepsilon )$ can be approximated by
its asymptotic behavior. Posteriorly, the integral can be expanded in the
parameter $\varepsilon .$ This procedure leads to the approximations given
in Eqs. (\ref{AjusteDelta1}) and (\ref{AjusteDelta2}).

The parameters of Eq. (\ref{AjusteDelta2}) are%
\begin{equation}
a=\frac{(1-\mu )(1-2\mu )}{4\mu -3}\{3-2H[2(1-\mu )]\},
\end{equation}%
where $H[x]=\gamma +\psi (x-1),$ where $\gamma $ is the Euler constant and $%
\psi (x)$ is the digamma function,%
\begin{equation}
b=\frac{2}{4\mu -3}[(1-\mu )(2\mu -1)],
\end{equation}%
and%
\begin{equation}
c=\frac{(1-2\mu )^{2}}{(4\mu -3)^{2}}\frac{\Gamma (1+\lambda )}{\Gamma (4\mu
-2+\lambda )},
\end{equation}%
while%
\begin{eqnarray}
d &=&\frac{-1}{(4\mu -3)}\frac{\Gamma (1+\lambda )}{\Gamma (4\mu -2+\lambda )%
} \\
&&\times \Big{[}\frac{1}{4\mu -1}+\frac{2^{1-4\mu }}{\sqrt{\pi }}\Gamma (%
\frac{1}{2}-2\mu )\Gamma (2\mu )\Big{]}.  \notag
\end{eqnarray}

\end{document}